\documentclass[aps,prl,twocolumn,amsmath,amssymb,superscriptaddress,longbibliography]{revtex4-1}
\usepackage[english]{babel}
\usepackage{xcolor}
\usepackage{amssymb} 
\usepackage{dcolumn}
\usepackage{bm}
\usepackage{graphicx}
\usepackage{amsmath}
\usepackage{braket}

\usepackage{graphicx}        

\graphicspath{{pict/}{}}

\usepackage[normalem]{ulem}

\usepackage{dcolumn}
\usepackage{bm}
\usepackage[pdfstartview=FitH, CJKbookmarks=true, bookmarksnumbered=true, bookmarksopen=true, colorlinks=true, pdfborder=001, citecolor=blue, linkcolor=blue, urlcolor=blue, linktocpage=true] {hyperref}

\makeatletter

\begin{document}


\title{Many-Body Amplified Nonclassical Photon Emission in Cavity-Coupled Atomic Arrays}

\author{Jing Tang}
\affiliation{School of Physics and Optoelectronic Engineering, Guangdong University of Technology, Guangzhou 510006, China}
\affiliation{Guangdong Provincial Key Laboratory of Sensing Physics and System Integration Applications, Guangdong University of Technology, Guangzhou, 510006, China}

\author{Yuangang Deng}
\email{dengyg3@mail.sysu.edu.cn}
\affiliation{Guangdong Provincial Key Laboratory of Quantum Metrology and Sensing $\&$ School of Physics and Astronomy, Sun Yat-Sen University, Zhuhai 519082, China}

\begin{abstract}
The generation of high-performance nonclassical light remains a cornerstone of quantum technologies, yet faces a fundamental trade-off between emission purity and brightness. Here, we demonstrate that cavity-mediated many-body spin-exchange interactions provide a route to overcome this constraint by collectively amplifying spectral anharmonicity. In a cavity-coupled atomic array with a programmable relative phase $\phi$, the resulting interference-interaction mechanism reshapes the dressed-state manifold and enables deterministic switching between distinct quantum emission regimes. For $\phi=0$, constructive interference yields high-purity single-photon emission with antibunching improved by four orders of magnitude while preserving strong photon flux. Conversely, for $\phi=\pi$, destructive interference creates a dark single-photon manifold, resonantly activating two-photon processes to produce bright and pure photon-pair bundles. Our work establishes interference-engineered many-body interactions as a scalable mechanism for on-demand quantum light generation and open a new avenue for harnessing collective many-body physics in quantum photonics.
\end{abstract}
\date{\today} 

\maketitle

{\em Introduction}.---Engineering light-matter interactions is central to quantum information and simulations~\cite{RevModPhys.80.885,RevModPhys.82.2313,RevModPhys.95.035002}. Recent advances in ultracold quantum gases coupled to optical cavities provide powerful platforms to explore strongly correlated many-body physics under controllable light-matter coupling~\cite{RevModPhys.85.553,RevModPhys.91.025005,mivehvar2021cavity}. In particular, infinite-range cavity-mediated spin-exchange interactions (SEI) have enabled the realization of diverse collective phenomena, including superradiant lasing~\cite{norcia2018cavity}, nonequilibrium quantum dynamics~\cite{muniz2020exploring,luo2024momentum,PhysRevLett.134.113403,you2025spin}, and dark superradiance in polar molecules~\cite{wang2025dark}. In parallel, optical tweezer arrays of neutral atoms and polar molecules provide a highly controllable setting that combines single-particle addressability with programmable long-range interactions~\cite{ebadi2021quantum,barredo2016atom,anderegg2019optical}, enabling controlled studies of collective radiation~\cite{Leonard2017,PhysRevLett.131.253603,PhysRevLett.130.173601,PhysRevLett.133.243401}, symmetry-protected phases~\cite{de2019observation,yue2025observing}, quantum-enhanced sensing~\cite{marciniak2022optimal,PhysRevLett.123.260505,PRXQuantum.5.010311}, and deterministic entanglement generation~\cite{wilk2010entanglement,holland2023demand,bao2023dipolar,schine2022long}.

\begin{figure}[ht]
\includegraphics[width=0.95\columnwidth]{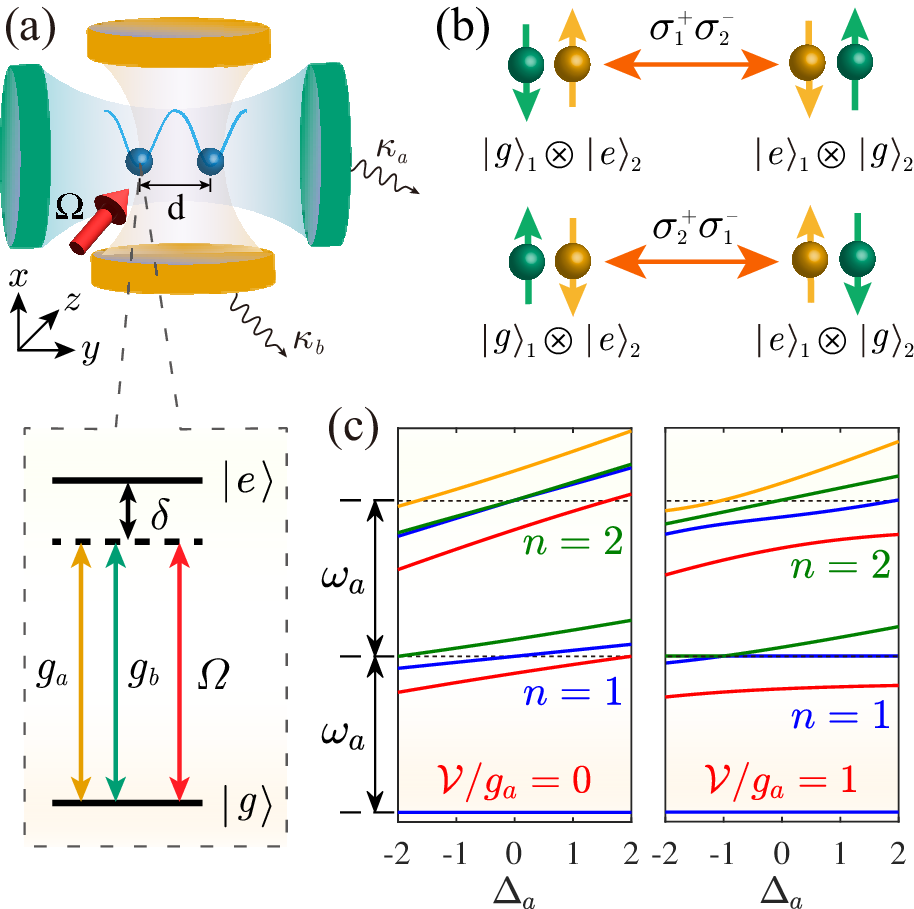}
\caption{(a) Schematic of cavity-coupled atomic arrays and (b) cavity-mediated SEI. (c) $\Delta_a$-dependent anharmonic energy spectrum.}
\label{model1}
\end{figure} 

Meanwhile, generating nonclassical light from antibunched single photons~\cite{birnbaum2005photon,dayan2008photon,PhysRevLett.118.133604,PhysRevLett.134.183601,PhysRevLett.134.013602} to strongly correlated multiphoton states~\cite{munoz2014emitters,liu2023deterministic,PhysRevLett.133.043601}, constitutes a central resource for quantum networks~\cite{ritter2012elementary,redjem2023all}, quantum communication~\cite{Duan01, Kimble08}, and quantum-enhanced metrology ~\cite{niemietz2021nondestructive, Giovannetti2011,RevModPhys.90.045005,RevModPhys.90.035005}. Despite significant progress, high-fidelity multiphoton emission typically relies on strong intrinsic optical nonlinearities, which remain challenging to achieve and control in  experiments~\cite{PhysRevA.110.030101, chakram2022multimode, carl2023phases}. Harnessing collective long-range interactions to engineer effective optical nonlinearities and controllable photon statistics remains largely unexplored, offering a promising path to novel many-body photonics states~\cite{defenu2024out,young2024observing}. 

In this Letter, we propose an experimentally accessible scheme that enables controllable switch between single-photon and two-photon bundle emission in two orthogonal cavities coupled to reconfigurable atomic arrays. By adiabatically eliminating the far-detuned auxiliary cavity, we engineer a tunable infinite-range SEI among atoms. The interplay of SEI and phase-programmable interference gives rise to two distinct regimes of nonclassical light emission. For $\phi=0$, constructive interference cooperates with SEI-induced spectral anharmonicity, yielding strong photon blockade, with single-photon purity enhanced by four orders of magnitude compared with the case without SEI. In contrast, for $\phi = \pi$, destructive interference creates a dark single-photon manifold that suppresses single-photon excitation while resonantly activating two-photon channels and inhibiting higher-order processes even at moderate SEI strength. This mechanism produces high-purity two-photon bundles under experimentally accessible parameters~\cite{norcia2018cavity,muniz2020exploring,luo2024momentum,PhysRevLett.134.113403}.

Remarkably, we further reveal a pronounced alternation of spin-spin correlations: positive (negative) transverse and negative (positive) longitudinal correlations accompany single-photon (multiphoton) emission, providing a direct diagnostic for distinguishing nonclassical states. Unlike previously explored routes to photon bundles based on Mollow physics~\cite{munoz2014emitters}, parametric down-conversion~\cite{PhysRevLett.117.203602}, multiphoton resonance~\cite{deng2021motional}, or parity-symmetry-protected mechanism~\cite{PhysRevLett.127.073602}, our scheme relies on interference-engineered many-body interactions rather than intrinsic optical anharmonicity. This phase- and SEI-controlled mechanism is naturally scalable to larger emitter networks, offering a reconfigurable route toward on-demand multiphoton sources for quantum technologies~\cite{neuzner2016interference,wang2025scalable, PhysRevResearch.6.033247}. 

{\em Cavity-coupled atomic array with SEI}.---We consider two identical atoms trapped in a one-dimensional optical tweezer array along $y$-axis and coupled to two orthogonal optical cavities with wave vector $k$, as illustrated in Fig.~\ref{model1}(a). Each atom is modeled as a two-level system consisting of a ground state $|g\rangle$ and a long-lived excited state $|e\rangle$, which may correspond to dipole-forbidden $^1S_0\!\leftrightarrow\!{}^3P_1$ transition of alkaline-earth atoms, featuring a narrow spontaneous decay $\gamma=(2\pi)7.5$ kHz~\cite{PhysRevLett.118.263601}. The  optical transition $|g\rangle\leftrightarrow|e\rangle$ couples to two cavity modes with single-atom cavity coupling $g_a$ ($g_b$) and cavity decay $\kappa_a$ ($\kappa_b$), respectively. Cavity mode A is chosen to be resonant with atomic transition, while cavity B operates in far-detuned dispersive regime, with light-cavity detunings $\Delta_a$ and $\Delta_b$. In addition, the atoms are driven by a transverse pump field with Rabi frequency $\Omega$.

In the far-dispersive limit $|g_b/\Delta_b|\gg 1$, the auxiliary cavity can be adiabatically eliminated, resulting in an effective Hamiltonian for atoms coupled to cavity A 
\begin{align}
{\cal {\hat{H}}}_1 & = \Delta_a\hat{a}^\dag\hat{a} + \frac{1}{2}\sum_{j=1}^2 (\delta \hat{\sigma}_j^z + \Omega \hat{\sigma}_j^x) +  {\cal V}\sum_{j\neq k}\hat{\sigma}_j^+\hat{\sigma}_k^- \nonumber \\
&+  g_a \hat{a}^\dag (\hat{\sigma}_1^- +  \cos \phi\,\hat{\sigma}_2^-) +  {\rm H.c.},
\label{mHam} 
\end{align}
where $\hat{a}$ denotes annihilation operators of cavity, $\hat{\sigma}_j^{x,y,z}$ are Pauli operators for atom $j$, and $\phi=kd$ is the tunable phase determined by interatomic separation $d$. The cavity-mediated SEI is ${\cal V}=-g_b^2\Delta_b/(\Delta_b^2+\kappa^2_b)$ and $\delta=\delta_c+{\cal V}/2$ with $\delta_c$ being the bare atom-pump detuning. The emerged SEI realizes a one-axis twisting interaction [Fig.~\ref{model1}(b)], opening a many-body gap that suppresses detrimental changes of collective spin induced by single-particle dephasing~\cite{luo2024momentum,PhysRevLett.134.113403}. 

\begin{figure}[ht]
\includegraphics[width=0.98\columnwidth]{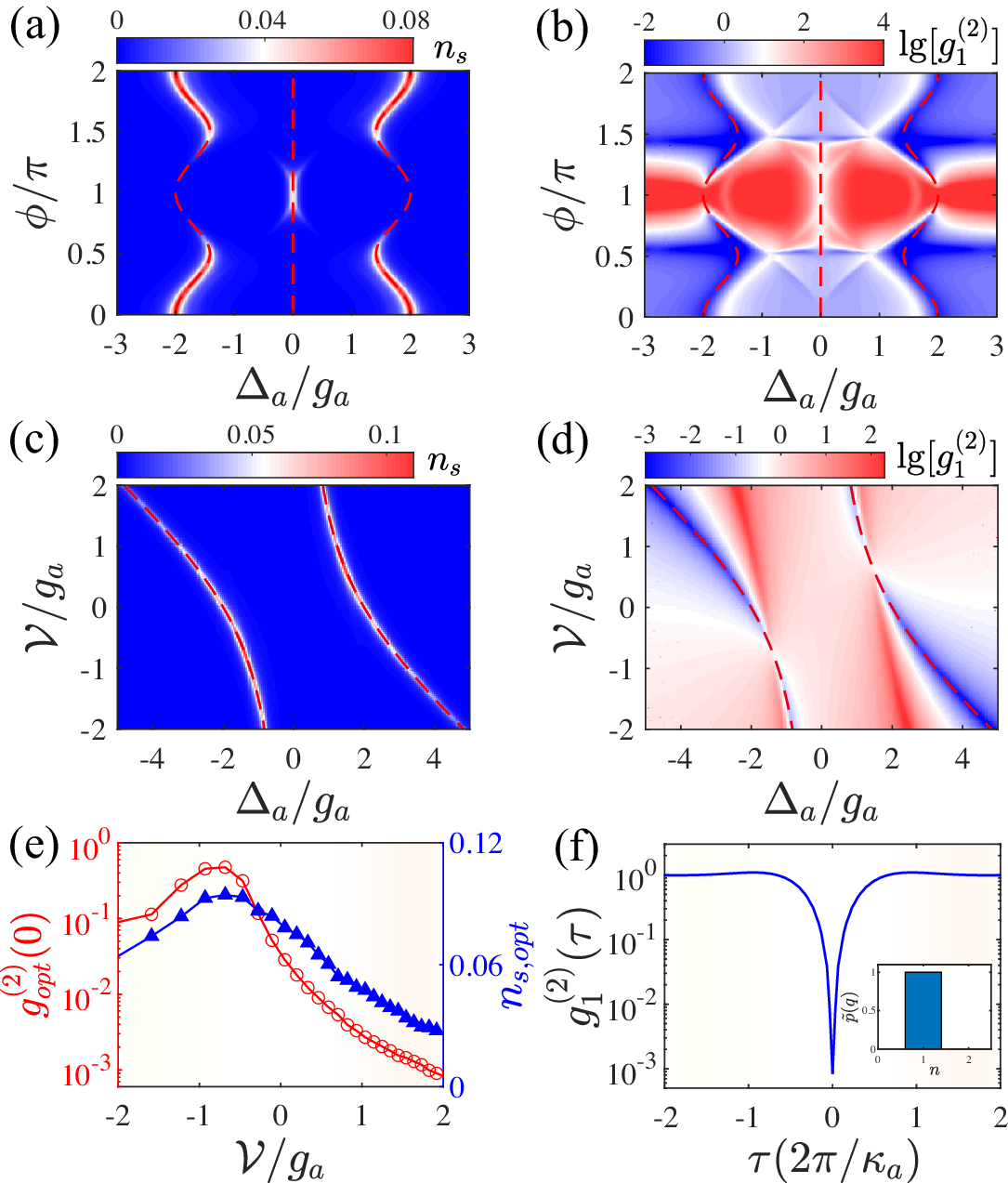}
\caption{(a) $n_s$ and (b) $g_1^{(2)}(0)$ as functions of $\Delta_a$ and $\phi$ for ${\cal{V}}=0$ and $\delta/\Delta_a=1/2$.  (c) $n_s$ and (d) $g_1^{(2)}(0)$ as functions of  $\Delta_a$ and ${\cal{V}}$ for $\phi/\pi =0$. (e) Optimal $g_{\mathrm{opt}}^{(2)}(0)$ (red line) and corresponding $n_{s,\mathrm{opt}}$ (blue line) versus ${\cal V}$ at $\Delta_a=-{\cal V}-\sqrt{{\cal V}^2+4g_a^2}$. (f) $\tau$ dependence $g_1^{(2)}(\tau)$ at  ${\cal{V}}/g_a=2$. The inset in (f) shows typical distribution $\tilde{p}(q)$ of single photon state.}
\label{phi}
\end{figure}

Figure~\ref{model1}(c) illustrates $\Delta_a$-dependent anharmonic energy spectrum of cavity-coupled atomic array for different $n$-photon dressed states. Three (four) helicity branches appear in the single- (two-) excitation manifold. For ${\cal V}=0$, two degenerate dark states appear at $\Delta_a=0$ for photon number $n \geq 2$, giving rise to three single-photon resonance points. In contrast, for ${\cal V}/g_a=1$ and $\phi=\pi$, the degeneracy of dark states is lifted, leaving a single dark-state protected by destructive quantum interference. The resulting cooperation between SEI and phase-controlled interference markedly enhances spectral anharmonicity, providing deterministic control of nonclassical photon emission.

{\em Results}.---The quantum statistics of emitted nonclassical light are obtained by solving full master equation including complete dissipations. In numerical simulation, we choose experimentally realistic parameters: $g_a=(2\pi)120$ kHz and $\kappa_a=(2\pi)15$ kHz~\cite{941q-5sdq,kongkhambut2022observation,PhysRevLett.118.263601}, $g_b=(2\pi)5$ MHz and $\kappa_b=(2\pi)0.5$ MHz~\cite{Leonard2017,PhysRevLett.131.253603, PhysRevLett.130.173601}, and a weak transverse drive $\Omega/\kappa_a=0.2$. By tuning the auxiliary cavity into far-dispersive regime $|\Delta_B|>(2\pi) 100$ MHz, cavity-mediated SEI can be continuously controlled over the range ${\cal V}/g_a\in[-2,2]$. The associated inelastic decay induced by eliminated cavity is $\gamma_e=\kappa_b g_b^2/(\Delta_b^2+\kappa_b^2)< (2\pi)1.25 ~{\rm kHz}$~\cite{norcia2018cavity}. All parameters are well within reach of state-of-the-art cavity-QED experiments. 

In the absence of SEI (${\cal V}=0$), $n$-photon dressed spectrum ($n \geq 2$) hosts \emph{two degenerate dark states} for $\phi=0$ ($-$) and $\pi$ ($+$), given by
 \begin{eqnarray}\label{mdark}
|\psi_{n,A_{\mp}}\rangle &\!=\!\!& \frac{1}{\sqrt{2}}\left(|n\!-1,g,e\rangle {\mp} |n\!-\!1,e,g\rangle\right), \\
|\psi_{n,S_{\mp}}\rangle&\!=\!\!&\sqrt\frac{n-1}{{2n-1}}|n,g,g \rangle \mp \sqrt\frac{n}{{2n-1}}|n-2,e,e\rangle,\nonumber
\end{eqnarray}
with $\delta/\Delta_a=1/2$. These dark states originate from distinct physical mechanisms. The antisymmetric state $|\psi_{n,A_-}\rangle$ is protected by exchange symmetry, which suppresses single-photon excitation through destructive interference. By contrast, the symmetric counterpart $|\psi_{n,A_+}\rangle$ under particle exchange, also exhibits vanishing single-photon occupation. The second class of dark state $|\psi_{n,S_\mp}\rangle$ arises from interference between excitation pathways involving different photon numbers and is decoupled from single-photon manifold. Consequently, single-photon channel is suppressed while resonant two-photon transitions remain allowed, leading to purely two-photon bundle emission via super-Rabi oscillations $\left| 2, g,g\right\rangle \leftrightarrow \left| 0, e,e \right\rangle$~\cite{zhao2025tunable}. This mechanism is fundamentally distinct from earlier interference schemes that lead to strictly vanishing photon emission~\cite{PhysRevLett.130.173601}. 

Figures~\ref{phi}(a) and~\ref{phi}(b) show the steady-state photon number $n_s$ and zero-delay second-order correlation function $g_1^{(2)}(0)$ on $\Delta_a$-$\phi$ parameter plane for ${\cal V}=0$. Pronounced photon emission occurs at the single-photon resonances, including the sidebands $\Delta_a/g_a = \pm \sqrt{2(1+\cos^2\phi)}$ and middle branch $ \Delta_a/g_a = 0$. For two bright polariton branches, $n_s$ decreases monotonically with increasing $\phi$ and is completely suppressed at $\phi=\pi$. At $\phi=0$, the emitted field exhibits strong photon antibunching, with $g_1^{(2)}(0) \simeq 2.7 \times 10^{-2}$, reflecting constructive to destructive quantum interference in single-photon excitation channel.

In contrast, $n_s$ for middle branch increases monotonically with $\phi$. At $\phi=\pi$, destructive interference selectively blocks single-photon excitation, allowing correlated multiphoton processes to dominate. Both dark states $|\psi_{n,A_+}\rangle$ and $|\psi_{n,S_+}\rangle$ in Eq.~(\ref{mdark}) are symmetric under particle exchange, leaving higher-order multiphoton excitations accessible. Remarkably, photon statistics exhibit a sharp transition to strong bunching with $g_1^{(2)}(0) \simeq 10$ [Fig.~\ref{phi}(b)], signaling multiphoton emission enabled by single-photon dark state formation. This phase-controlled antibunching-bunching transition is fundamentally distinct from conventional photon blockade~\cite{birnbaum2005photon,dayan2008photon,PhysRevLett.118.133604,PhysRevLett.134.183601,PhysRevLett.134.013602}, as it originates from tunable quantum interference rather than purely from spectral anharmonicity.

\begin{figure}[ht]
\includegraphics[width=0.93\columnwidth]{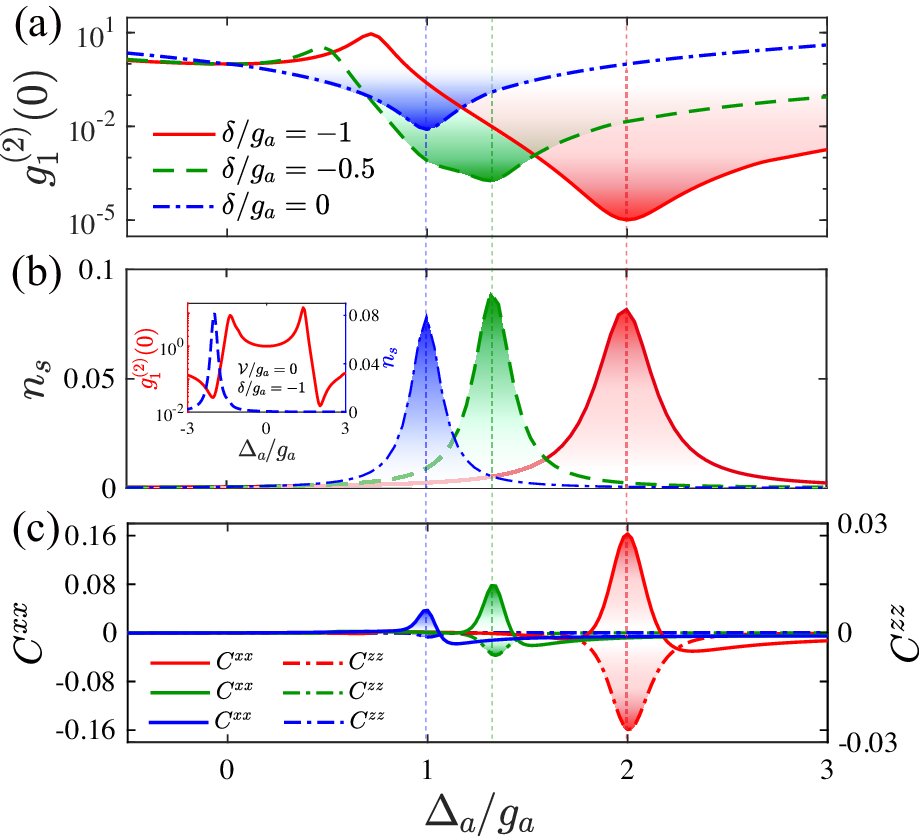}
\caption{(a) $g_1^{(2)}(0)$, (b) $n_s$, and (c) $C^{\mu\mu}$ as a function of $\Delta_a$ for different $\delta$ with ${\cal{V}} =2$. The inset in (b) illustrates the corresponding $g_1^{(2)}(0)$ and $n_s$ for ${\cal{V}} =0$ and $\delta/g_a=-1$.}
\label{delta_Delta}
\end{figure}

\begin{figure*}[ht]
\includegraphics[width=1.93\columnwidth]{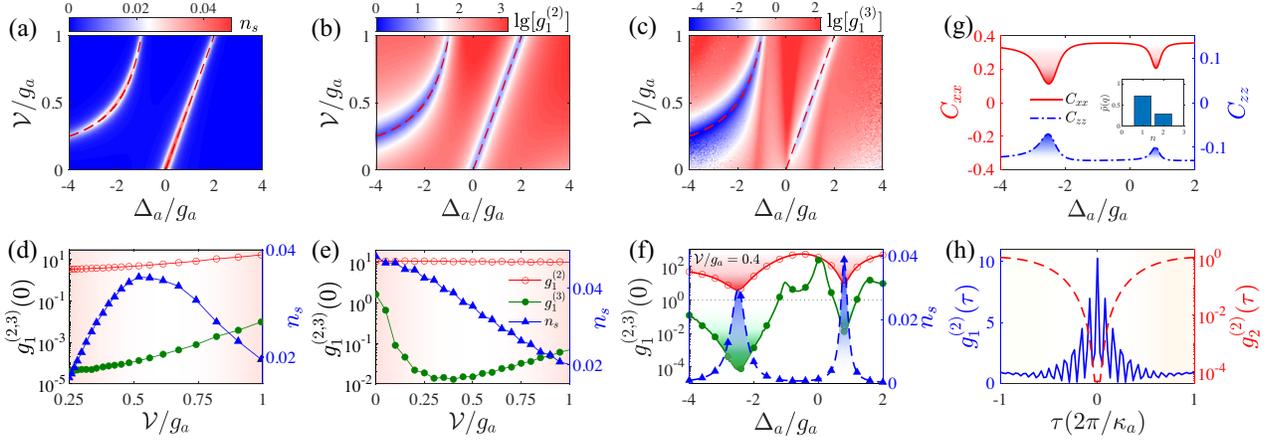}
\caption{(a-c) shows $n_s$, $g_1^{(2)}(0)$, and  $g_1^{(3)}(0)$ in the ${\cal{V}}$-$\Delta_a$ parameter plane for $\delta/ {\cal{V}} =-1$ and $\phi/\pi=1$.  (d,e) $g_1^{(2,3)}(0)$ and $n_s$  as a function of ${\cal V}$ evaluated along the trajectories $\Delta_a=-g_a^2/{\cal V}$  (d) and $\Delta_a=2{\cal V}$  (e), respectively. (f) $\Delta_a$ dependent $g_1^{(2,3)}(0)$ and $n_s$, and (g) the corresponding spin correlations $C^{\mu\mu}$ for a fixed interaction strength ${\cal{V}}/g_a=0.4$. (h) $\tau$ dependent $g_1^{(2)}(\tau)$ (blue line) and $g_2^{(2)}(\tau)$ (red line) for  $\Delta_a=2{\cal V}=0.8 g_a$. }
\label{V_Deltac_pi}
\end{figure*} 

{\em Single-photon emission}.---Introducing SEI  strongly reshapes the anharmonic energy spectrum and substantially enhances the optical nonlinearity [Fig.~\ref{model1}(c)]. Figures~\ref{phi}(c) and \ref{phi}(d) show $n_s$ and $g_1^{(2)}(0)$ as functions of ${\cal{V}}$ and $\Delta_a$. Because the antisymmetric state $|\psi_{n,A_{-}}\rangle$ is completely decoupled for the middle branch, photon emission is dominated by the two bright states. Accordingly, $g_1^{(2)}(0)$ exhibits a pronounced dependence on both magnitude and sign of ${\cal V}$. The resulting red-blue-detuning asymmetry is consistent with the analytically predicted vacuum Rabi splitting $\Delta_a=-{\cal V}\pm \sqrt{{\cal V}^2+4g_a^2}$. 

To quantify the optimal single-photon performance, we extract $g_{opt}^{(2)}(0)$=min[$g_1^{(2)}(\Delta_a)$] and the corresponding $n_{s,opt}$ at the same detuning as a function of ${\cal V}$ [Fig.~\ref{phi}(e)]. Remarkably, $g_{opt}^{(2)}(0)$ remains well below unity across the entire interaction range, indicating robust photon antibunching at sideband resonances. Compared with ${\cal V}=0$, the antibunching is enhanced by two orders of magnitude, reaching $g_{opt}^{(2)}(0)=8\times10^{-4}$ at ${\cal V}/g_a=2$, albeit with a reduced photon population $n_{s,{\rm opt}}$. 

The single-photon nature is further confirmed by time-dependent second-order correlation function $g_1^{(2)}(\tau)$  [Fig.~\ref{phi}(f)]. The condition $g_1^{(2)}(\tau)>g_1^{(2)}(0)$ demonstrates pronounced antibunching, while $g_1^{(2)}(\tau)$ relaxes to unity on a timescale $\tau=1/\kappa_a=10.6$ ms. This antibunching lifetime is enhanced by nearly four-orders of magnitude compared with single-atom photon blockade~\cite{birnbaum2005photon}. Consistently, steady-state photon-number distribution $\tilde{p}(q)=q p(q)/n_s$ with $p(q)=\langle q|\hat{a}^\dagger\hat{a}|q\rangle$ is strongly peaked at single- photon state, yielding $\tilde{p}(1)=0.99998$ and strongly suppressed higher-order occupations. These results confirm that the observed strong antibunching originates from SEI-enhanced spectral anharmonicity.

In contrast to Jaynes-Cummings model, photon statistics in the presence of SEI (${\cal V}\neq0$) can be efficiently tuned by varying detuning $\delta$. For single-photon excitation, the vacuum Rabi splitting $\Delta_a=2g_a^2/(\delta+{\cal V})$ increases rapidly as $|\delta+{\cal V}|$ approaches zero, directly reflecting the amplification of spectral anharmonicity. High-quality single-photon emission therefore emerges from SEI-amplified optical nonlinearity. To elucidate this behavior, we plot $g_{1}^{(2)}(0)$ and $n_s$ as a function of $\Delta_a$ for different values of $\delta$ with fixing ${\cal{V}} =2$ [Figs.~\ref{delta_Delta}(a) and \ref{delta_Delta}(b)]. As $\Delta_a$ increases, photon antibunching is rapidly enhanced, reflecting the growth of vacuum Rabi splitting induced by SEI-amplified spectral anharmonicity. Remarkably, photon number remains nearly unchanged even deep in the strong photon-blockade regime with $g_{1}^{(2)}(0)=1\times10^{-5}$. Compared to the noninteracting case (${\cal V}=0$) [inset of Fig.~\ref{delta_Delta}(b)], single-photon purity is improved by more than three orders of magnitude while maintaining a comparable photon occupation ($n_s \simeq 0.08$).  

The coexistence of extremely strong antibunching \emph{without sacrificing brightness}, demonstrates that a high quality single-photon source can be engineered through SEI-amplified optical nonlinearity. This behavior stands in sharp contrast to conventional photon blockade schemes~\cite{birnbaum2005photon,dayan2008photon,PhysRevLett.118.133604,PhysRevLett.134.183601,PhysRevLett.134.013602}, where enhanced antibunching typically occurs at cost of strongly reduced photon flux. Notably, such enhancement is entirely absent for ${\cal V}=0$, underscoring the essential role of long-range SEI in amplifying spectral anharmonicity.

Further physical insight is provided by examining connected spin correlations $C^{\mu\mu}=\langle \hat{\sigma}^\mu_{1}\hat{\sigma}^\mu_{2}\rangle - \langle \hat{\sigma}^\mu_{1}\rangle \langle\hat{\sigma}^\mu_{2}\rangle$ with $\mu=x, z$~\cite{PhysRevLett.133.106901}. Figure~\ref{delta_Delta}(c) shows the transverse correlation $C^{xx}$ (solid line) and longitudinal correlation $C^{zz}$ (dashed-dotted line) as a function of $\Delta_a$ for different $\delta$. As $\delta$ decreases, $C^{xx}$ becomes strongly positive while $C^{zz}$ develops a large negative, indicating enhanced transverse spin coherence accompanied by anti-ferromagnetic correlations along $z$-axis. These correlations directly shape photon statistics: positive $C^{xx}$ enhances collective emission within single-excitation manifold, boosting photon flux, whereas negative $C^{zz}$ suppresses double excitation, enforcing strong antibunching. The resulting synergy between collective emission and reinforced photon blockade enables the simultaneous realization of high brightness and near-ideal single-photon purity. Remarkably, these results identify the detuning $\delta$ as a powerful control knob for engineering nonclassical light in the presence of long-range SEI. Strong antibunching can be achieved while maintaining an almost constant photon flux, paving the way toward high-fidelity single-photon sources.

{\em Two-photon bundle emission}.---We now turn to the regime $\phi=\pi$, where single-photon emission is completely suppressed by destructive quantum interference. Crucially, this interference-engineered spectrum offers a microscopic switching mechanism between single-photon ($\phi=0$) and two-photon bundle ($\phi=\pi$) emission. Figures~\ref{V_Deltac_pi}(a-c) display phase-space maps of $n_s$, $g_1^{(2)}(0)$, and $g_1^{(3)}(0)$ as functions of ${\cal V}$ and  $\Delta_a$ at $\delta/ {\cal{V}} =-1$. At two-photon resonances $\Delta_a=-g_a^2/{\cal V}$ and $2{\cal V}$, two distinct branches associated  with pronounced $n_s$ are clearly resolved. In sharp contrast to single-photon emission, photon statistics satisfy $g_1^{(2)}(0) \gg 1$ and $g^{(3)}(0)\ll 1$, signaling emerged photon-pair bunching accompanied by three-photon blockade.

To quantify the quality of photon-pair emission, we extract the optimal values of $\log[g_1^{(2,3)}(0)]$ and $n_s$ as a function of  ${\cal{V}}$ for two resonant branches, as shown in Figs.~\ref{V_Deltac_pi}(d) and \ref{V_Deltac_pi}(e). Along blue sideband $\Delta_a=-g_a^2/{\cal V}$, both $\log[g_1^{(2)}(0)]$ and $\log[g_1^{(3)}(0)]$ decrease monotonically as ${\cal V}$ reduced. Notably, photon-pair bunching $g_1^{(2)}(0)>1$ and ultrastrong three-photon blockade $g_1^{(3)}(0)\sim10^{-4}$ persist even for a weak SEI (${\cal V}/g_a<0.48$). This strong suppression of higher-order photon emission and the nonmonotonic behavior of $n_s$, demonstrate the generation of high-purity photon pairs. For red sideband $\Delta_a=2 {\cal V}$, increasing ${\cal{V}} $ produces a shallow rise in $g_1^{(2)}(0)$ and a nonmonotonic reduction of $g_1^{(3)}(0)$, reaching a minimum $g_1^{(3)}(0)=10^{-2}$ while maintaining a sizable photon number ($n_s=0.04$). 

These features originate from phase-controlled quantum interference at $\phi/\pi=1$, which renders single-excitation manifold dark while leaving two-excitation manifold resonantly accessible. The inclusion of SEI provides a powerful knob to tune both position and strength of two-photon resonances. Figure~\ref{V_Deltac_pi}(f) displays $\Delta_a$-dependence of photon statistics for a weak interaction ${\cal{V}} /g_a=0.4$, unambiguously demonstrating high-purity and high-brightness photon-pairs generation on both sidebands. By contrast, two-photon emission at middle branch vanishes when ${\cal{V}} =0$. Achieving comparable performance without SEI would require an unrealistically small dissipation ratio $\gamma/\kappa_a \ll 1$. Notably, even moderate SEI suppresses $g_1^{(3)}(0)$ by more than four orders of magnitude compared to noninteracting configuration, highlighting its essential role in amplifying spectral anharmonicity.

Further insight is obtained by examining the connected spin correlation $C^{\mu\mu}$ as a function of $\Delta_a$ [Fig.~\ref{V_Deltac_pi}(g)]. Different from single-photon regime, two-photon resonance exhibits a distinct correlation signature: transverse component $C^{xx}$ is strongly suppressed, while longitudinal component $C^{zz}$ becomes significantly less negative. This behavior reflects the inhibition of single-excitation processes and enhanced participation of double excitations. Consistently, the time-dependent correlation functions $g_1^{(2)}(\tau)$ and $g_2^{(2)}(\tau)$ [Fig.~\ref{V_Deltac_pi}(h)] satisfy $g_1^{(2)}(0)>g_1^{(2)}(\tau)$ and $g_2^{(2)}(\tau)>g_2^{(2)}(0)$, confirming that photons are emitted in strongly correlated two-photon bundles. The photon-number distribution $\tilde{p}(q)$ is correspondingly dominated by two-photon events, with multiphoton contributions suppressed to $\sum_{q>2}\tilde{p}(q)=7\times 10^{-6}$, demonstrating the realization of high-fidelity  two-photon sources.

{\em Conclusion}.--We have proposed an experimentally feasible scheme in which a controllable atomic phase, combined with cavity-mediated SEI enables deterministic switching between single-photon and two-photon bundle emission. The SEI-enhanced spectral anharmonicity selectively addresses distinct multiphoton manifolds, yielding either strong photon antibunching or bright correlated photon-pair emission. These two emission regimes are distinguished by opposite transverse and longitudinal spin-correlation signatures, providing a direct and versatile diagnostic of nonclassical photon states via atomic spin-correlation measurements. Our work establishes interference-engineered many-body interactions as a general platform for high-purity, high-brightness multiphoton sources, with potential applications in precision metrology and fundamental physics~\cite{wilk2010entanglement,holland2023demand,bao2023dipolar,schine2022long}.

{\em Acknowledgments}.---This work was supported by the National Natural Science Foundation of China (Grant No.12374365, Grant No. 12274473, and Grant No. 12135018), Quantum Science and Technology-National Science and Technology Major Project (Grant No.2025ZD0300400), Guangdong Provincial Quantum Science Strategic Initiative (Grant No. GDZX2505001), and Guangdong University of Technology SPOE Seed Foundation (SF2024111504).

\onecolumngrid

\begin{center}
  \textbf{\large End Matter}
\end{center}
\vspace{0.5em}

\twocolumngrid

\appendix  
\setcounter{equation}{0}
\renewcommand{\theequation}{A\arabic{equation}}

{\em Quantum statistics}.--To characterize quantum correlations of the emitted nonclassical light, we describe the nonequilibrium dynamics of the system by a Lindblad master equation that incorporates all relevant dissipation channels,
 \begin{equation}\label{master equation}%
{ \frac{d\rho}{dt}}= -i [\hat{\cal H}_1, {\rho}] + {\kappa}_a \mathcal {\cal{D}}[\hat{a}]\rho + ({\gamma}+\gamma_e)\sum_{j=1}^2 \mathcal
{\cal{D}}[\hat{\sigma}_{j}^-]\rho,
\end{equation}
where $\rho$ is the density matrix of cavity-coupled atomic array and $\mathcal {D}[\hat{o}]\rho=\hat{o} {\rho} \hat{o}^\dag - (\hat{o}^\dag \hat{o}{\rho} + {\rho}\hat{o}^\dag \hat{o})/2$ denotes standard Lindblad dissipator. An additional effective atomic decay rate $\gamma_e=\kappa_b g_b^2/(\Delta_b^2+\kappa_b^2)$ arises from the adiabatic elimination of the auxiliary cavity mode~\cite{norcia2018cavity}.

The generalized $k$th-order correlation function is then defined as~\cite{munoz2014emitters, del2012theory}
\begin{align}
g_n^{(k)}(\tau_1,\dots,\tau_k) = \frac{\left\langle \prod_{i=1}^k \left[\hat{a}^{\dagger}(\tau_i)\right]^n \prod_{i=1}^k \left[\hat{a}(\tau_i)\right]^n \right\rangle}{\prod_{i=1}^k \left\langle \left[\hat{a}^{\dagger}(\tau_i)\right]^n \left[\hat{a}(\tau_i)\right]^n \right\rangle},
\label{g220}
\end{align}
with $\tau_1 \leq \dots \leq \tau_k$. This quantity provides a unified description of nonclassical emission processes ranging from single photons to correlated $n$-photon bundles. For single-photon emission, the conditions $g^{(2)}_1(0)<1$ and $g_1^{(2)}(0)<g_1^{(2)}(\tau)$ signify sub-Poissonian statistics and photon antibunching. In contrast, an $n$-photon emission ($n\ge2$) requires $g^{(n)}(0) > 1$ together with $g^{(n+1)}(0) < 1$, indicating the suppression of higher-order excitations~\cite{PhysRevLett.118.133604}. For genuine $n$-photon bundle emission, the criteria $g_1^{(2)}(0)>g_1^{(2)}(\tau)$ and $g_n^{(2)}(0)<g_n^{(2)}(\tau)$ ensure photon bunching within each bundle and antibunching between separated bundles~\cite{munoz2014emitters, deng2021motional}.

\setcounter{equation}{0}
\renewcommand{\theequation}{B\arabic{equation}}
{\em Energy spectrum}.---To elucidate the dressed-state structure underlying interference-controlled photon emission, we analyze the energy spectrum of cavity-coupled atomic array subject to the combined effects of cavity-mediated SEI and phase-programmable quantum interference. Neglecting the weak coherent driving field in Hamiltonian (\ref{mHam}), the total excitation number $\hat{N}=\hat{a}^\dag\hat{a} +\sum_{j}\hat{\sigma}_j^+\hat{\sigma}_j^-$ conserved. Consequently, the accessible Hilbert space decomposes into independent excitation manifolds labeled by photon number $n$. Within a fixed-$n$ manifold, the accessible basis states are $\Psi=[|n-1, g, e \rangle,|n-1, e, g \rangle, |n-2,e, e \rangle, |n, g, g \rangle]^T$, where the kets denote photon number and the internal states of the two atoms. The stationary Schr$\ddot{o}$dinger equation ${\hat{\cal H}}\Psi={\cal M}\Psi$ then yields the effective Hamiltonian matrix ${\mathcal {M}}$,
\begin{align}   \label{matrix1}
\left(\!\begin{array}{@{}c@{}c@{}c@{}c@{}} 
\xi & {\cal V}& \sqrt{n-1}g_a & \sqrt{n}g_a \cos \phi \\
 {\cal V} & \xi & ~~\sqrt{n-1}g_a \cos \phi & \sqrt{n}g_a \\
\sqrt{n-1}g_a & \sqrt{n-1}g_a \cos \phi~~ & (n-2)\Delta_a+2\delta & 0 \\
\sqrt{n}g_a \cos \phi & \sqrt{n}g_a & 0 & n\Delta_a \\
\end{array}\!\right) 
\end{align}
where we have introduced the shorthand $\xi= (n-1)\Delta_a+\delta$.

Introducing the symmetric and antisymmetric collective atomic states $|\pm\rangle =  \bigl(|g,e\rangle \pm |e,g\rangle
\bigr)/{\sqrt2}$, the SEI term $\mathcal{V} (\hat{\sigma}_1^+\hat{\sigma}_2^- + \hat{\sigma}_2^+\hat{\sigma}_1^-)$ becomes block-diagonal, producing energy shifts $\pm \mathcal{V}$ for the symmetric and antisymmetric states, respectively. Proceeding in this collective basis, the accessible Hilbert space within a fixed photon-number manifold $n$ is spanned by the Fock states vector $\Psi=[|n\!-\!1,+\rangle,|n\!-\!1,-\rangle,|n\!-\!2,e,e\rangle,|n,g,g\rangle]^T$. Then the Schr$\ddot{o}$dinger equation ${\hat{\cal H}}\Psi={\cal M}\Psi$  yields  the relevant matrix ${\mathcal {M}}$,
\begin{equation}   \label{matrix}
\begin{pmatrix}
\xi_+ & 0 & \sqrt{n-1} \, g_+ & \sqrt{n} \, g_+ \\
0 & \xi_- & \sqrt{n-1} \, g_- & -\sqrt{n} \, g_- \\
\sqrt{n-1} \, g_+ & \sqrt{n-1} \, g_- & (n-2)\Delta_a+2\delta & 0 \\
\sqrt{n} \, g_+ & -\sqrt{n} \, g_- & 0 &  n\Delta_a
\end{pmatrix},
\end{equation}
with energy shift $\xi_\pm = (n-1)\Delta_a+\delta \pm{\cal V}$ and phase-dependent coupling $g_\pm = g_a(1 \pm \cos\phi)/\sqrt{2}$. Notably, the combined control of SEI ${\cal V}$ and phase $\phi$ thus provides a powerful and tunable mechanism for engineering photon statistics, by shaping quantum interference between distinct atom cavity excitation pathways and modifying many-body excitation gaps, thereby suppressing Doppler-induced dephasing~\cite{luo2024momentum,PhysRevLett.134.113403}.

For nonzero SEI (${\cal V}\neq 0$), the effective Hamiltonian (\ref{matrix}) becomes block-diagonal at  $\phi = 0$, while the antisymmetric state $|n-1,-\rangle$ is completely decoupled from cavity field due to $g_- = 0$. Focus on the single- excitation manifold ($n=1$), two eigenenergies associated with the sideband branches are given by
\begin{equation}
\begin{aligned}
E_{1,\pm}&=\frac{\Delta_a+\delta+{\cal V}}{2} \pm \frac{1}{2} \sqrt{(\delta+{\cal V}-\Delta_a)^2 + 8g_a^2},
\end{aligned}
\end{equation}
which correspond to collective vacuum Rabi splitting $\Delta_a=2g_a^2/(\delta+{\cal V})$. Notably, the effective anharmonicity of the spectrum is strongly enhanced as $|\delta+{\cal V}|\rightarrow 0$, where the level spacing becomes highly nonlinear. This regime supports high-quality single-photon emission, characterized by strong photon antibunching and large steady-state intracavity photon population. 

In stark contrast, for $\phi=\pi$ and ${\cal V}=-\delta$, destructive quantum interference gives rise to dark state $|\psi_{n,A_{+}}\rangle=|n-1,+\rangle$. In this case, both single- and two-photon resonances occur at the same detuning, $\Delta_a=-g_a^2/{\cal V}$, with a collective vacuum Rabi splitting $\sqrt{(2{\cal V}-g_a^2/{\cal V})^2+8g_a^2}$. The corresponding eigenstates are
\begin{align}\label{two1}
|\psi_{2,1}\rangle&=\frac{\sqrt{2}{g_a\cal V}|1,- \rangle +g_a^2|0,e,e \rangle -{\sqrt{2}{\cal V}^2}|2,gg \rangle}{\sqrt{g_a^4+2{g_a^2\cal V}^2+4{\cal V}^4}}, \nonumber\\
|\psi_{1,1}\rangle&=\frac{g_a|0,- \rangle -{\sqrt{2}{\cal V}}|1,gg \rangle}{\sqrt{g_a^2+2{\cal V}^2}},
\end{align}
This spectral structure reveals the coexistence of resonant single- and two-photon excitation channels. Moreover, an additional two-photon resonance appears at $\Delta_a=2{\cal V}$, associated with the eigenstate
\begin{align}\label{two2}
|\psi_{2,2}\rangle&=\frac{\sqrt{2}{\cal V}|1,- \rangle +{g_a}|0,e,e \rangle +\sqrt{2}g_a|2,gg \rangle}{\sqrt{3g_a^2+{\cal V}^2}},
\end{align}
which explicitly corresponds to a correlated two-photon bundle state. Taken together, this interference-controlled energy spectrum  provides the microscopic mechanism for quantum switching between single- and two-photon bundle emission. Importantly, the presence of SEI not only enables this switching but also substantially enhances the purity of both single- and two-photon quantum light sources.

\end{document}